\documentclass[pre,preprint,letterpaper,showpacs,superscriptaddress]{revtex4}
\usepackage{epsfig}
\usepackage{graphicx}
\usepackage{amsmath}
\usepackage{amssymb}
\begin{document}
%\draft
\title{Trapping mechanism in overdamped ratchets with quenched noise}
\author{D.G. Zarlenga}
\author{H.A. Larrondo} \thanks{CONICET Researcher},
\author{C.M. Arizmendi}
\affiliation{Departamento de F\'{\i}sica, Facultad de
Ingenier\'{\i}a, Universidad Nacional de Mar del Plata,\\  Av. J.B.
Justo 4302, 7600 Mar del Plata, Argentina\\}
\author{Fereydoon Family}
\affiliation{Department of Physics, Emory University, Atlanta, GA
30322,  USA}
\date{\today }
%\maketitle
\begin{abstract}
A trapping mechanism is observed and proposed as the origin of the
anomalous behavior recently discovered in transport properties of
overdamped ratchets subject to external oscillatory drive in the
presence of quenched noise. In particular, this mechanism is shown
to appear whenever the quenched disorder strength is greater than a
threshold value.  The minimum disorder strength required for the
existence of traps is determined by studying the trap structure in a
disorder configuration space.  An approximation to the trapping
probability density function in a disordered region of finite length
included in an otherwise perfect ratchet lattice is obtained. The
mean velocity of the particles and the diffusion coefficient are
found to have a non-monotonic dependence on the quenched noise
strength due to the presence of the traps.

\end{abstract}

\maketitle
\section{Introduction}
The existence of chaotic behavior, which is the seemingly random
complex motion observed in deterministic nonlinear systems is now
well established.  In particular, many approaches have been
developed for characterizing and understanding the nature of chaotic
motion \cite{strogatz}.

In addition to chaotic behavior, it has also been shown that
deterministic systems can exhibit anomalous transport and strange
kinetics \cite{scher, shlesinger, klafter, lichtenberg}. In analogy
with stochastic processes, in the case of normal diffusion
\cite{geisel, black}, the mean square displacement $<x^2>$ is
proportional to time $t$ ($<x^2>\sim t$), while in the case of
strange kinetics \cite{shlesinger, klafter, barkai}, $<x^2>\sim
t^\gamma$, with $\gamma>2$ for enhanced diffusion and $1<\gamma<2$
for dispersive motion. The mean square displacement can also have a
logarithmic dependence on time, corresponding to $\gamma=0$
\cite{marinari, krug}. Strange kinetics as well as diffusive motion
have been observed in both deterministic nonlinear systems
\cite{morgado, mallic, lenzi, korabel, kunz} as well as thermal
ratchets \cite{popescu}.

In this paper we concentrate on the dynamics of a deterministic
thermal ratchet in the presence of a driving force. It has recently
been shown \cite{popescu} that quenched disorder induces a normal
diffusive kinetics in addition to the drift due to the external
drive. Moreover this diffusive motion is enhanced by higher values
of the quenched disorder. If the quenched disorder has long-range
spatial correlations, diffusion becomes anomalous, and both the
correlation degree and the amount of quenched disorder can enhance
the anomalous diffusive transport \cite{popescuchin}.
%
%
%%%%%% begin of changed paragraph %%%%%%% %%%%%%% %%%%%%% %%%%%%% %%%%%%% %%%%%%% %%%%%%%
Anomalous transport has been found recently in overdamped systems
\cite{linder, reimann2}. In Linder et al \cite{linder} an anomalous
coherence is reported and P. Reimann et al \cite{reimann2} find
divergence on the diffusion coefficient. Although our system differs
from those previously reported due to the presence of quenched
disorder and driving force, the transport anomaly presents some
similarities that will be discussed below.
%%%%%%%%%% end of changed paragrapph %%%%%%% %%%%%%% %%%%%%% %%%%%%% %%%%%%% %%%%%%% %%%%%%%
%
%
While anomalous transport in quenched disorder ratchets was observed
for a range of values of the parameters, the mechanism leading to
this unusual behavior has not been investigated. Transport
properties of ratchets (for a recent review of ratchets, see
\cite{reimann}) is a topic of great current interest, due to the
possible application of these models for understanding such systems
as molecular motors \cite{Astumian97, AstumianHanggi02}, nanoscale
friction \cite{daikhin, daly, soren}, surface smoothening
\cite{baraba}, coupled Josephson junctions \cite{zapata}, as well as
mass separation and trapping at the microscale \cite{gorre, deren,
ertas, duke}. The fluctuations that produce the net transport are
usually associated with noise, but they may arise also in absence of
noise, with additive forcing, in overdamped deterministic systems
\cite{Barbi7}, overdamped quenched systems \cite{popescu} and in
underdamped ratchets \cite{jung, mateos, barbi, quenched2}.

The aim of this paper is to show that a trapping mechanism is
responsible for the observed dispersive anomalous transport in an
overdamped ratchet subject to an external oscillatory drive
\cite{motmol}. In particular, we show that this mechanism appears
when the quenched disorder strength is greater than a threshold
value.  The minimum disorder strength required for the existence of
traps is determined by studying the trap structure in a disorder
configuration space.  An approximation to the trapping probability
density function in a disordered region of finite length included in
an otherwise perfect ratchet lattice is obtained. We show that due
to this trapping mechanism, the mean velocity of the particles and
the diffusion coefficient have a non-monotonic dependence on the
quenched noise strength.

The outline of the paper is as follows: in section II we present the
single particle model, in section III we define the ensemble and the
cumulants, in section IV we present the trapping mechanism and
discuss its consequences in section V.  Conclusions are presented in
section VI.

\section{Model}
The motion of a single particle in an overdamped disordered media is
modeled by an overdamped ratchet subject to an external oscillatory
drive in the presence of a quenched noise, using the dynamical
equation:
\begin{equation}\label{overquen}
\gamma \dot{x}=R(x)+F(t)+G(x)
\end{equation}
where, $\gamma $ is the damping coefficient, $R(x)=-dU/dx$ is the
ratchet force, $F(t)$ is the time dependent external force and
$G(x)$ is the quenched disorder force.

The periodic, asymmetric, ratchet potential is modeled by the
equation:
\begin{equation}\label{potential}
U(x)=-\sin (x)-\frac{\mu }{2}\sin (2x).
\end{equation}%
with the spatial period $\lambda =2\pi$, as in previous works
\cite{popescu, jung, mateos, barbi, quenched2, larro}. The external
oscillatory force is given by:
\begin{equation}
F(t)=\Gamma \sin (\omega t) ,
\end{equation}
where $\Gamma $ and $\omega $ are the amplitude and the frequency of
the oscillations, respectively. The effects of the substrate
randomness is modeled by a quenched disorder term of the form:
\begin{equation}\label{noise}
G(x)=\alpha\sum_{i=-\infty}^{\infty} \xi (i)
[H(x-i\lambda)-H(x-(i+1)\lambda)] ,
\end{equation}
where the coefficient $\alpha \geq 0$ is the quenched disorder
strength, $H$ is the Heaviside function and $\xi (i)$ are
independent, uniformly distributed random numbers in $[-1,1]$.  The
extension to correlated disorder is straightforward.  The force
$G(x)$ is a piecewise constant force for every period of the ratchet
potential and gives a reasonably realistic representation of the
effects of the substrate.

In order to carry out a numerical solution of Eq. \ref{overquen} we
have carried out a fourth order fixed step Runge-Kutta method
\cite{numrec}. Since we are interested in the influence of $\Gamma$
and $\alpha$ on the transport properties of the system, the
remaining parameters are set to the following values, which were
used in previous works \cite{popescu, quenched2, motmol, sincro}:
\begin{equation}
\gamma=1;\mu=0.5;\omega=0.1.
\end{equation}
We have carried out numerical solutions of the evolution equation
using the following dimensionless variables:
\begin{itemize}
\item The dimensionless position
$\widetilde{x}=x/\lambda$, which gives the position of the particle
along the valleys of the ratchet potential.
\item The dimensionless velocity
$\widetilde{v}=v/v_{\omega }$, with $ v_{\omega }=\lambda /T$. The
mean value of $\widetilde{v}$ gives the transport velocity of a
particle along the ratchet. \item The discrete sequences obtained by
sampling $\widetilde{x}$ and $\widetilde{v}$ with a sampling period
$T_{sa}=T=2\pi/\omega$:
\begin{equation}
\widetilde{x}_{sa}=\widetilde{x}(kT)\quad ;\quad
\widetilde{v}_{sa}=\widetilde{v}(kT)\quad,
\end{equation}
with (k=0,1,2,...). Using these variables it is possible to detect
synchronization with the external driving force.
\end{itemize}

A typical trajectory of Eq. \ref{overquen} consists of an
oscillation superimposed on a directed transport motion with average
speed $<\widetilde{v}>$. The particular case of $<\widetilde{v}>=0$
indicates no transport along the ratchet.

\begin{figure}
\includegraphics
[width=14 cm]{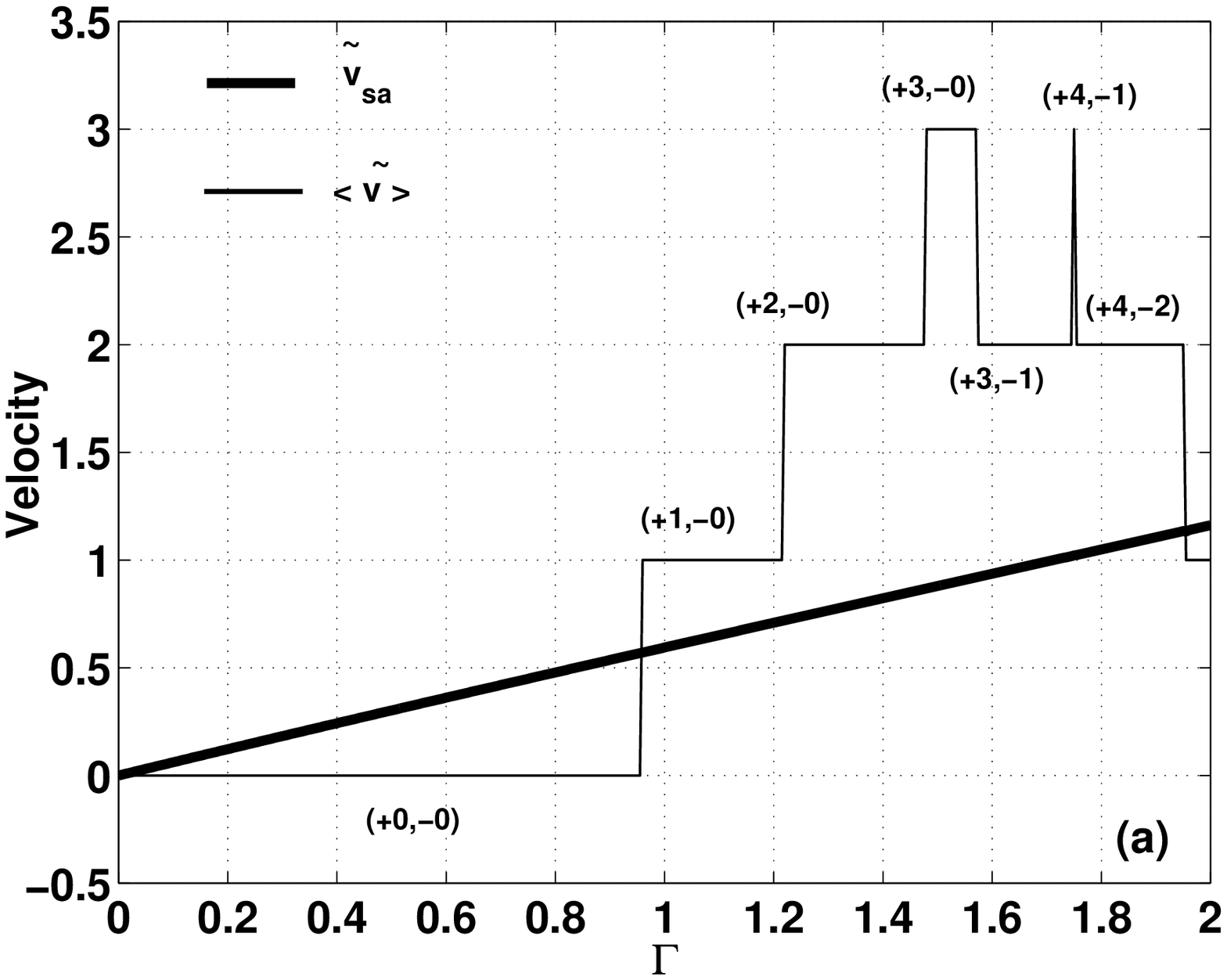}, \caption { Sampled velocity
$\widetilde{v}_{sa}$, and mean velocity $\langle \widetilde {v}
\rangle $ of a particle in a perfect lattice, as a function of
$\Gamma  $. The particle starts at $\widetilde{x}=0$.   Note the
jumps in $\langle \widetilde {v} \rangle $ at $\Gamma \simeq
0.96,1.22,1.47,1.57,1.75,1.95$, but $\widetilde {v}_{sa}$ has no
bifurcations in this range of $\Gamma $.  The label over each zone
indicates the number of valleys crossed by the particle, forward (+)
and backwards (-) in a period $T$.}\label{zonas}
\end{figure}

In a perfect lattice (i.e. $\alpha=0$) massless particles remain
synchronized over the entire $[\Gamma,\omega]$ parameter space
\cite{sincro}. The bifurcation diagrams of $\widetilde{v}_{sa}$ and
$<\widetilde{v}>$ as a function of $\Gamma$ are shown in
Fig.\ref{zonas} for $\Gamma $  in the range $[0,2]$. Fig.\ref{zonas}
shows that $\widetilde{v}_{sa}$ is a monotonic increasing function
of $ \Gamma $ and $<\widetilde{v}>$ is a stepped function with jumps
at specific $\Gamma $ values. The meaning of these jumps may be
understood by considering how the particle's position
$\widetilde{x}$ varies as a function of time $\widetilde{t}$. When
$\Gamma$ is below 0.96 the particle starts in a potential valley and
oscillates inside the valley in synchrony with the external driving
force, returning every $T$ to the same position inside the valley.
The particle also has the same velocity. This synchronism explains
why only one value of $\widetilde{v}_{sa}$ is obtained: at every
sampling time the velocity of the particle has the same value within
its oscillatory motion.

Over the region $\Gamma \in \left[ 0.96,1.22 \right] $, every
$\Gamma$ value has only one value for $\widetilde{v}_{sa}$ but now
$<\widetilde{v}>=1$, showing that the particle remains synchronized
with the external driving force but now it advances one spatial
period (one valley) during $T$. As $\Gamma $ further increases the
particle advances 2 valleys, and then 3 valleys during each $T$ (see
the labels $(+2,-0)$ and $(+3,-0)$ in Fig.\ref{zonas}), giving
$<\widetilde{v}>=2$ and $<\widetilde{v}>=3,$ respectively.
Furthermore it remains synchronized giving only one value of
$\widetilde{v}_{sa}$. If $\Gamma $ further increases,
$<\widetilde{v}>$ jumps to a lower value, because during the
positive half cycle the particle goes forward, crossing several
valleys, but it returns to one or more valleys during the negative
half cycle. This explains the labels in Fig. \ref{zonas}. The motion
of the particle remains synchronized with the external force through
the entire range $\left[ 0,2\right] $ as it is shown by the single
value of $\widetilde{v}_{sa}$.

\section{Collective Motion}
We have studied the evolution of an ensemble of noninteracting
particles, uniformly distributed over one whole potential valley.
The initial position of the particles is given by the particle
density function:
\begin{equation}\label{distriori}
\rho(x,0)=[H(\tilde{x}-\tilde{x}_{min})-H(\tilde{x}-\tilde{x}_{max})],
\end{equation}
with $\tilde{x}_{max}=\tilde{x}_{min}+1$.

The ensemble was allowed to evolve up to a time $\tilde{t}=1000$,
while the positions of the particles were obtained at times
$\tilde{t}_k=10k$ $(k=0, 1,...)$ and stored for further analysis. In
order to perform averages over the realizations of disorder a
different quenched disorder sequence was used for each trajectory .
In this way, the average over the trajectories also includes an
average over different realizations of the disorder.

To characterize the evolution of the packet, the first two
cumulants,
\begin{equation}
\widetilde{C}_{1}=\left\langle \widetilde{x}_{k}\right\rangle
;\text{ } \widetilde{C}_{2}=\left\langle
\widetilde{x}_{k}^{2}\right\rangle - \widetilde{C}_{1}^{2};
\end{equation}
and their temporal derivatives,
\begin{equation}
\ \langle V\rangle =\lim t\rightarrow \infty \left(
d\widetilde{C}_{1}/d \widetilde{t}\right) ;\text{\
}\widetilde{D}=\lim t\rightarrow \infty \left(
d\widetilde{C}_{2}/d\widetilde{t}\right)
\end{equation}
were evaluated at the sampling times as a function of time. Here
$\langle V\rangle$ is the mean velocity and $\widetilde{D}$ is the
diffusion coefficient. In all cases considered, it was verified that
all higher-order cumulants increase slower than $t^{n/2}$, ensuring
that $\rho(x,t)$ is asymptotically a Gaussian and can be determined
using the first two moments only.

\section{The trapping mechanism}

The superposition of the force $F(t)$, with zero temporal mean, and
the force $R(x)$ with a zero spatial mean value, allows the
particles to move at different speeds along the potential.
Consequently, in spite of the zero spatial mean value of the force
$R(x)$ produced by the ratchet potential:
\begin{equation}
\langle R\rangle
_{\widetilde{x}}=\frac{1}{N}\int\limits_{\widetilde{x_{1}}}^{
\widetilde{x_{1}}+N}R(\widetilde{x})d\widetilde{x}=\frac{1}{N}\left[U\left(
\widetilde{x} +N\right) -U\left( \widetilde{x}\right) \right]=0,
\end{equation}
the time-averaged mean value {\it felt} by the particles is not zero
and is given by:
\begin{equation}
\langle
R\rangle_{\widetilde{t}}=\frac{1}{N}\int\limits_{\widetilde{t_{1}}}^{
\widetilde{t_{1}}+1}R\left[ \widetilde{x}\left(\widetilde{t}\right)
\right] d\widetilde{t}.
\end{equation}
This is in fact the reason a sinusoidal driving force produces a
positive drift motion when it is combined with the ratchet
potential.

A significant consequence of the quenched disorder is the appearance
of a trapping mechanism, which arises only in the disordered case
($\alpha\neq 0$). Traps are a small number of contiguous valleys
with a negative time-averaged quenched-disorder mean-value, that
exactly compensate the positive time-averaged mean-value of the
ratchet potential force.  This trapping can be predicted from the
synchronization analysis of the perfect lattice case. As an example
let us consider the case $\Gamma=1.65 $. This $\Gamma$ corresponds
to the synchronization zone (+3,-1) in Fig.\ref{zonas}. When
$\alpha\neq 0$ a particle feels a force that is a combination of the
ratchet force plus a sinusoidal force with variable amplitude
$\Gamma_{eq}$ between $\Gamma-\alpha$ and $\Gamma+\alpha$. Thus,
disorder enables the particle reach different zones, as can be seen
in Fig. \ref{zonas}. For example, for $\alpha =0.1$, the available
regions for a particle are $(+3,-0)$ and $(+3,-1)$. Then the
possible values of $<\widetilde{v}>$ are a result of  the
combination of the positive and the negative terms: $+3-0=3$ and
$+3-1=2$. Thus, $\widetilde{v}$ is bounded between 2 and 3. For
$\alpha =\alpha_z\simeq 0.175$ the available regions become
$(+2,-0)$, $(+3,-0)$, $(+3,-1)$, $(+4,-1)$, and $(+4,-2)$. Then the
possible values of $<\widetilde{v}>$ are: $+4-0=4$, $+4-1=3$,
$+4-2=2$, $+3-0=3$, $+3-1=2$, $+3-2=1$, $+2-1=1$, $+2-2=0$. Since
zero is a possible value, then the particle can be localized or
trapped.  This corresponds to a particle going forward $2$ valleys
during the positive half cycle and going backwards $2$ valleys
during the negative half cycle. Consequently, the trapped particle
oscillates inside three valleys in synchrony with the external
driving force.  While this analysis is not exact, it provides a
reasonable explanation for both the minimal disorder strength and
the corresponding length $K$ in which particles can be trapped.

\begin{figure}
\includegraphics [width=8.5 cm]{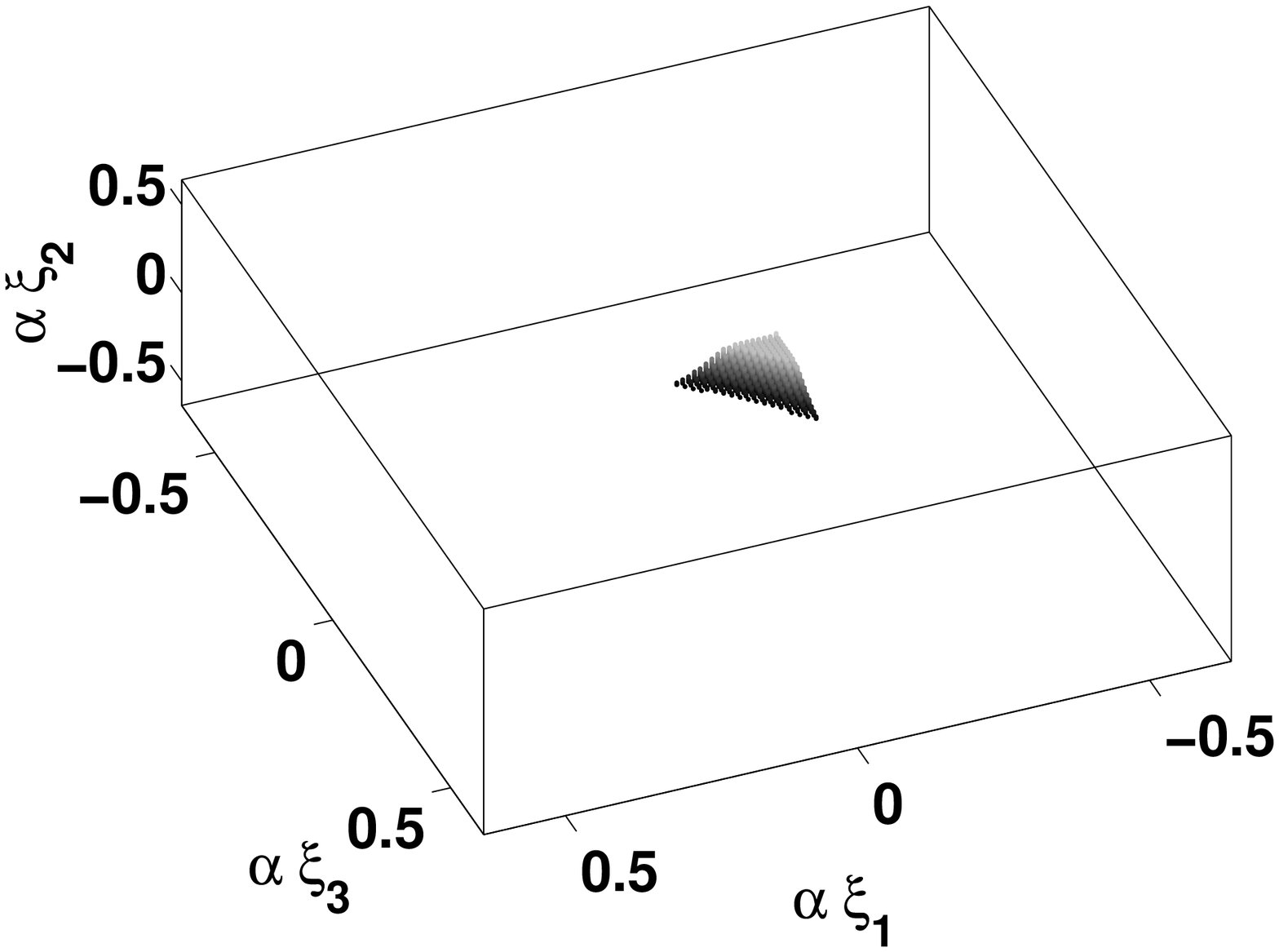}
\includegraphics [width=8.5 cm]{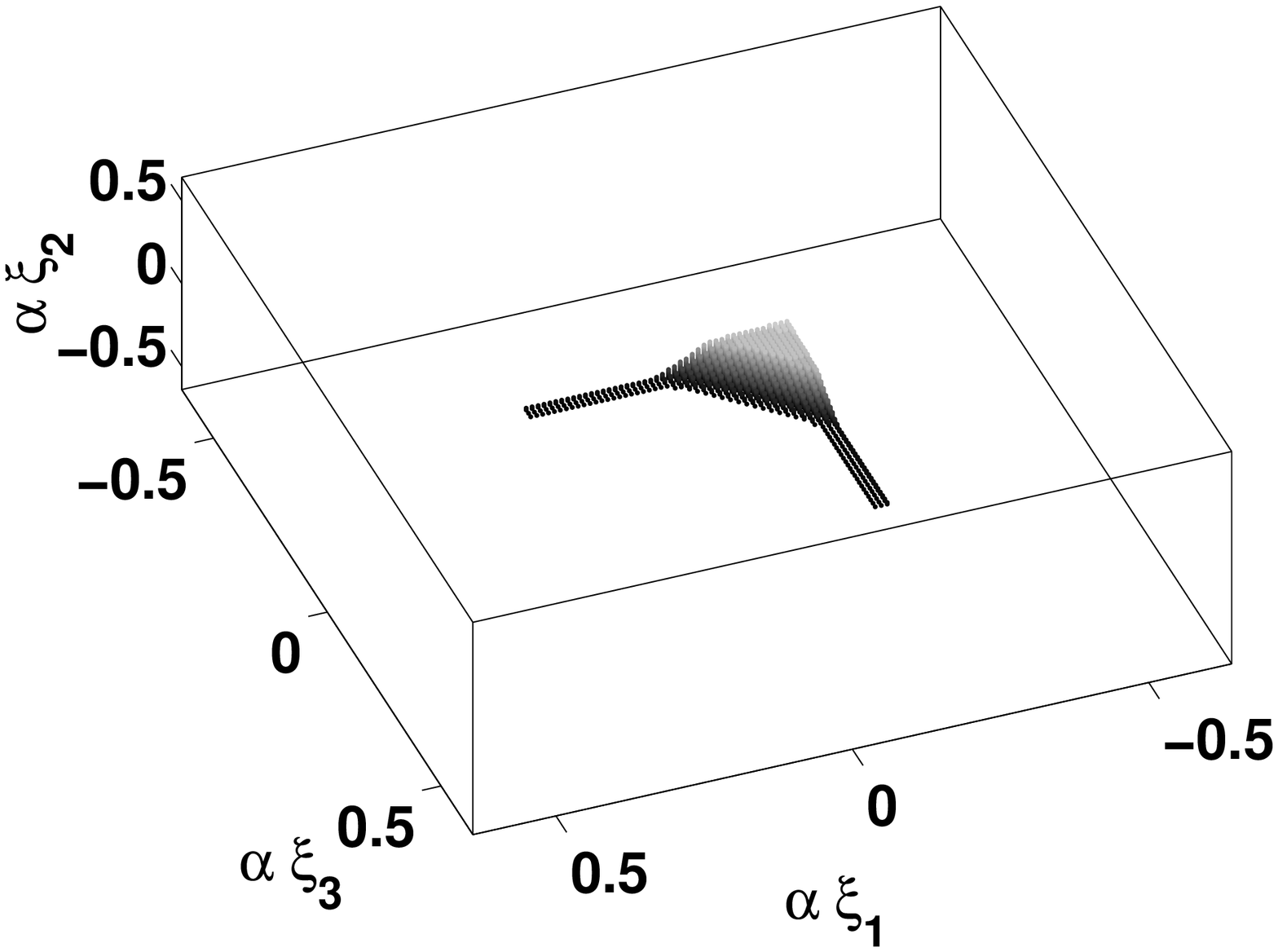}
\includegraphics [width=8.5 cm]{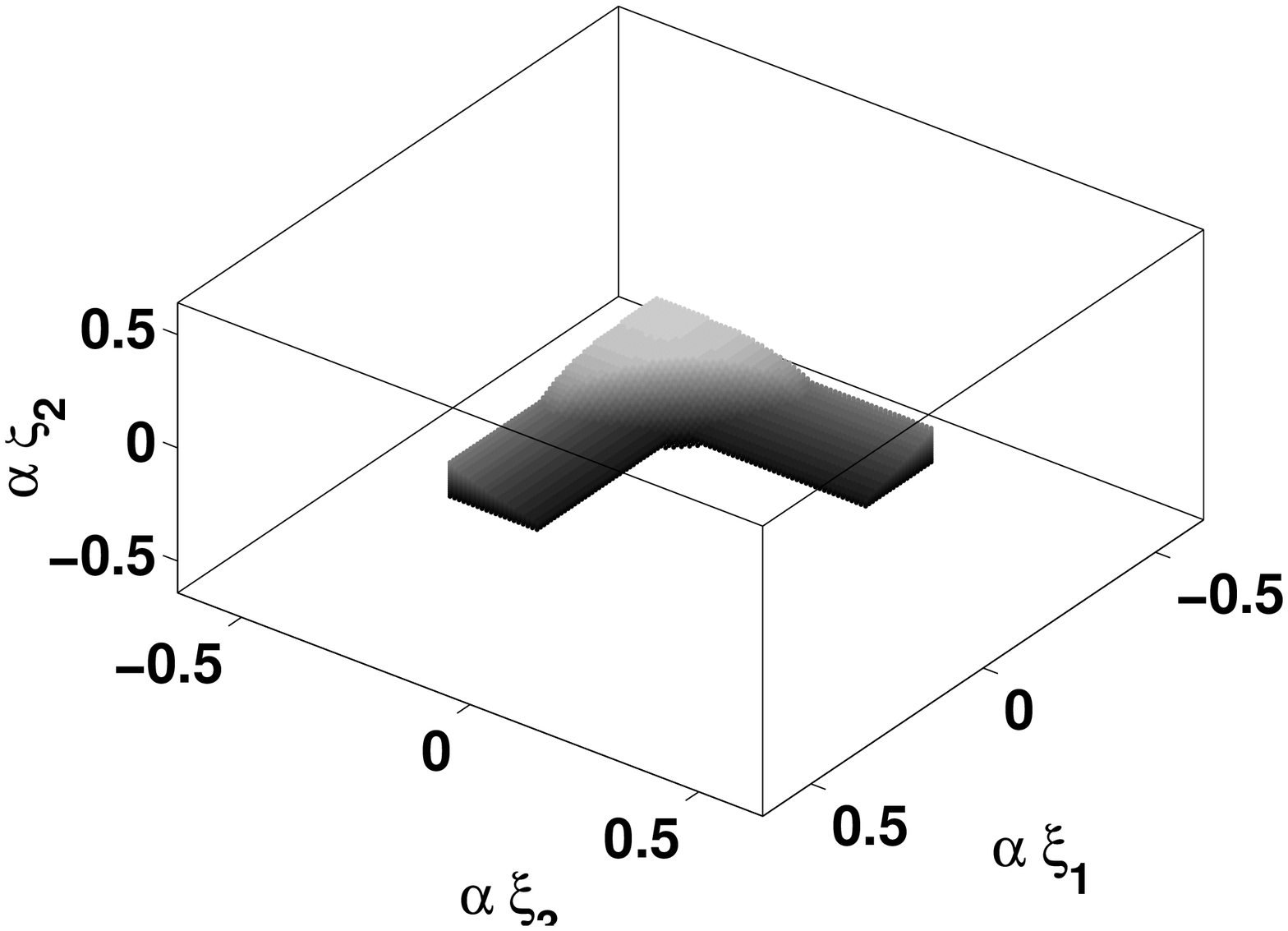}
\caption{Three dimensional disorder configuration space for $\Gamma
= 1.65$, $\omega=0.1$, $\gamma=1$ and $\mu=0.5$ as a function of
$\alpha$. The axes represent the quenched disorder in three
consecutive valleys. The black dots correspond to triplets acting as
traps. (a) The case $\alpha =0.20 \gtrapprox \beta_K $. Note the
small trapping region at the corner $\alpha \xi_1=\alpha
\xi_2=\alpha \xi_3=-\beta_K$, (b) The case $\alpha=0.22 \gtrapprox
\beta_{K-1}$.  Note the appearance of two dimensional arms at
$\alpha= \beta_{K-1}$; (c) The case $\alpha=0.30 > \beta_{K-1}$ }.
\label{volprob}
\end{figure}

In order to determine the trapping probability, we will first define
the quenched disorder forces $G(x)$ of Eq. \ref{overquen}, in one of
$K$ consecutive valleys, as the coordinate of a $K$-dimensional
disorder configuration space. Possible combinations of disorder are
studied in this space and each combination is classified either as a
\textit{trap} or a \textit{non-trap}. For example for $\Gamma=1.65$
$K=3$; then all possible combinations of three consecutive valleys
were studied.  The results are shown in a $K=3$ disorder
configuration space in Fig. \ref{volprob}. Note that the trapping
region in this space changes shape as $\alpha$ increases from 0.2 in
Fig. \ref{volprob}(a) to 0.3 in Fig. \ref{volprob}(c). For low
values of $\alpha$ there are no traps at all. At a critical value of
$\alpha\cong\beta_K=\beta_3$ the first trap appears (see
Fig.\ref{volprob}a). Let us call this trap
 the \textit{basic trap} as
it consists of $K$ consecutive valleys with equal disorder strength
(for the case $\Gamma=1,65$, $K=3$ and $\beta_3\cong0.136$).

Note that the volume of the trapping configuration region further
increases with increasing $\alpha$ (see Fig. \ref{volprob}b). For
$\alpha=\beta_{K-1}=\beta_2\cong0.21$ (see Fig. \ref{volprob}c), two
\textit{arms} appear, corresponding to traps of only $K-1$
consecutive valleys (for the case $\Gamma=1.65$, $K-1=2$). If
$\alpha$ is further increased, there exists a higher value
$\beta_{K-2}=\beta_1$  over which traps of only one valley appear.
Note that the minimum $\alpha$ for which traps appear is $\beta_K$

When $K=2$, the disorder configuration space is two-dimensional.
Traps appear when $\alpha>\beta_{K}=\beta_{2}$. In the interval
$\beta_{2}<\alpha<\beta_{1}$, the probability space (that
corresponds to trapping events) gradually mutates from a single
triangular shape into a square as $\alpha$ grows. When
$\alpha>\beta_{1}$, two narrow, rectangle-shaped arms appear,
because it is possible now for a particle to get trapped in just one
valley. We note that, in Fig. \ref{volprob}, in which a $K=3$ case
is considered, the $\beta_{1}$ value is outside the plotting range.
If $\beta_{1}$  had been included in the plot three rectangular
prisms would have appeared, one for each random variable. For
instance, the prism that corresponds to valley number $1$ would have
occupied the volume:

\begin{equation}\label{volu}
\begin{pmatrix}-\alpha<\alpha\xi_1<-\beta_1\\
-\alpha<\alpha\xi_2<\alpha\\
-\alpha<\alpha\xi_2<\alpha\\
\end{pmatrix}
\end{equation}.

For each $\alpha$ value the cumulative probability of trapping in a
$K$-length trap, called $p(\alpha,K)$, is evaluated as the ratio
between the volume of the trapping region and $(2\alpha)^K$.

When a disordered region of finite length $L>K$ is considered the
cumulative probability of trapping is approximately given by:
\begin{eqnarray}\label{proba}
  \nonumber P(\alpha,K,L)=1-Q(\alpha,K,L),\\
Q(\alpha,K,L)\cong\left(1-p(\alpha,K)\right)^{L-K+1}.
\end{eqnarray}
The approximation used to obtain Eq. \ref{proba} is based on the
assumption that the events of actual trapping of the particle in the
neighborhood of the  $K$-length traps are independent.

\begin{figure}
\includegraphics
[width=14 cm]{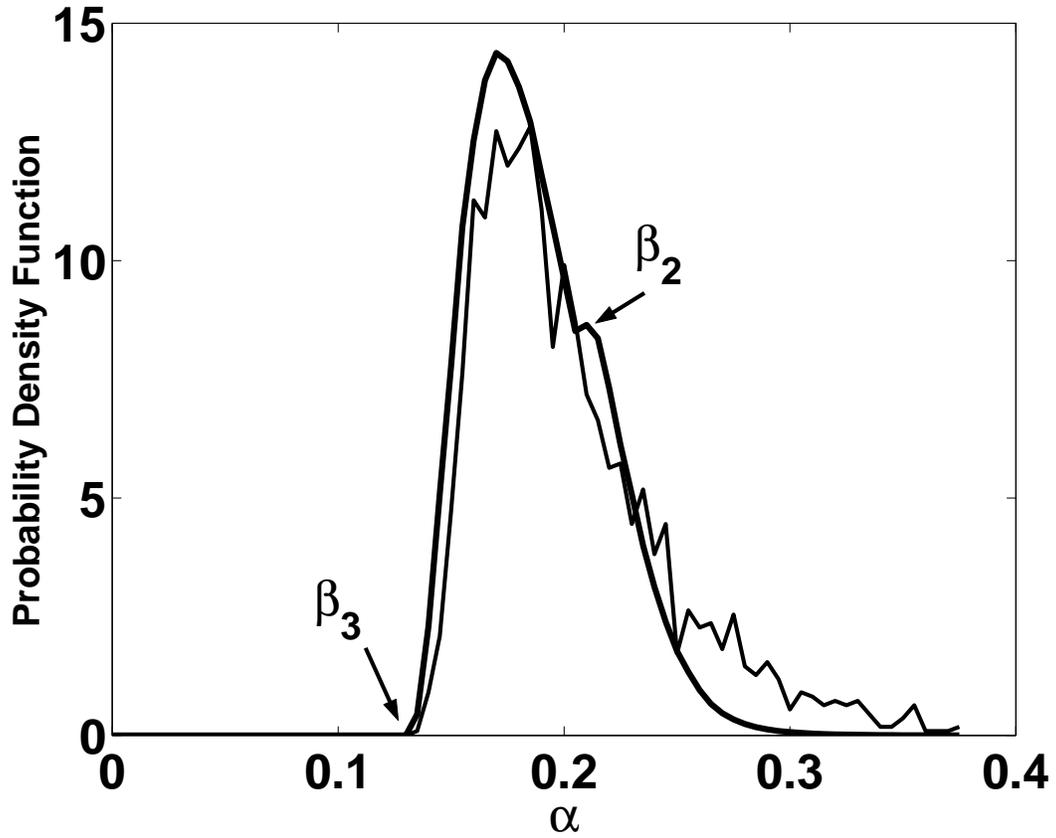} \caption {The probability density
function $dP(\alpha,K,L)/d\alpha$ from eq. \ref{proba} (bold curve)
and  the function $f(\alpha,L)$ from Eq. \ref{DprobaL} (thin curve)
for $2200$ massless particles traversing a $L=50$ disordered region,
are plotted as a function of $\alpha$, for $\Gamma =
1.65$,$\gamma=1$, $\omega=0.1$ and $\mu=0.5$. }\label{probafig}
\end{figure}

Let $F(\alpha,L)$ be the fraction of particles traversing the length
$L$ of the disordered region.  We define $f(\alpha,L)$ by the
relation:
\begin{equation}\label{DprobaL}
f(\alpha,L)=d\left[1-F(\alpha,L)\right]/d\alpha.
\end{equation}
In Fig. \ref{probafig}, $f(\alpha,L)$ is compared with
$dP(\alpha,K,L)/d\alpha$ with the values of $P(\alpha,K,L)$ obtained
from Eq. \ref{proba}. The good agreement between the two results
confirms that the independent events approximation used in Eq.
\ref{proba} is clearly valid for $\alpha$ values below
$\beta_{K-1}$.  Note that the local maximum of both curves in Fig.
\ref{probafig} occurs at $\beta_{K-1}\simeq0.21$, corresponding to
the value at which the "arms" begin to appear in the
disorder-configuration space (see Fig. \ref{volprob}b). We note that
the value $\beta_{K-2}\simeq0.5720$, where $K-2$ valley length traps
appear, falls outside the range of the plotted values.

As the disorder region length $L$ grows, the shape of Figure
\ref{probafig} becomes thinner but its left end is still at
$\alpha=\beta_{K}$ (in this case, $\alpha=\beta_{3}$). If $L$ tends
to infinity the probability that a particle finds a $K$-length trap
tends to unity when $\alpha=\beta_{K}$. Then, the shape of figure
\ref{probafig} becomes a Dirac delta function, which is also
predicted by Eqs. \ref{proba} and \ref{DprobaL}.

We have carried out similar studies for many other  values of
$\Gamma$ in the range $[0,2]$. We have found that in most of the
cases studied, the correlation effects are not important and Eq.
\ref{proba} is accurate enough for determining the probability
density function $P(\alpha,K,L)$. There are a few $\Gamma$ values,
however, for which the correlation effects cannot be easily
neglected, since $\beta_{K} \simeq \beta_{K-1}$ and the "arms" in
the disorder configuration space (like the one in figure
\ref{volprob}) appear early on.

As seen in Fig. 4, Eq. \ref{proba} is not accurate when $\alpha
>> \beta_{K}$. This is because the correlations cannot be neglected
in this region. We propose another method, which we call the {\it
Conditional-Probability Method} or CPM, that takes the correlations
into account. This method is always accurate provided that
$\alpha>>\beta_{1}$ but for some $\Gamma$ values it also works well
in the entire $\alpha>\beta_{1}$ range.

Calculations based on CPM involve the following steps: 1)
Approximate the probability space volume occupied by trapping events
(both black and gray dots in Fig. 3), by a polyhedron in which the
faces are perpendicular to the coordinate axes. The
$\alpha>\beta_{1}$ case must be considered and rectangle prisms are
the main volume component. The smaller volume components are
optionally considered, if the method's  range of valid $\alpha$
values is to be increased. If $K=2$ we have a probability space
enclosed by a surface with faces that are perpendicular to the
coordinate axes. 2) Write the volume equations of the non-trapping
events, that is the volume outside the polyhedron, as the union of
the set of volumes. The probability of having no traps is
proportional to this volume. For example, let us consider the
$\Gamma=1.35$ case for which $K=2$. The non-trapping probability
$Q(\alpha,K,L)= Q(\alpha,2,2)$ agrees approximately with:
\begin{equation}\label{Q2}
Q(\alpha,2,2)(4\alpha^2)=\begin{pmatrix}-\beta_1<\alpha\xi_1<\alpha\\-\beta_1<\alpha\xi_2<\alpha
\end{pmatrix}.
\end{equation}
If the smaller surface of trapping events approximately given by
\begin{equation}
-\beta_{1}<\alpha\xi_{1}<-\beta_{2},
-\beta_{1}<\alpha\xi_{2}<-\beta_{2}
\end{equation} is taken into
account we can instead write:

\begin{multline}\label{Q2a}
Q(\alpha,2,2)(4\alpha^2)=\left[\begin{pmatrix}-\beta_2<\alpha\xi_1<\alpha\\
                          -\beta_2<\alpha\xi_2<\alpha\\
    \end{pmatrix}\bigcup\begin{pmatrix}
-\beta_2<\alpha\xi_1<\alpha\\
-\beta_1<\alpha\xi_2<-\beta_2\\
\end{pmatrix}\bigcup\begin{pmatrix}
-\beta_1<\alpha\xi_1<-\beta_2\\
-\beta_2<\alpha\xi_2<\alpha\\
\end{pmatrix}\right].
\end{multline}

3) Increase $L$, sequentially, in single steps, and find the new
volume as the intersection of the previous volumes. That is,
$Q(\alpha,K,L-1).(2\alpha)^{(L-1)}$, expanded for all
$\alpha\xi_{L}$ and the original $K$-dimensional volume expanded for
all $\alpha\xi_{1}$, $\alpha\xi_{2}$, ...., $\alpha\xi_{L-K}$
values. Continuing with this procedure and using the most accurate
volume equation, we can write the $L=3$ volume in the following way:

\begin{multline}\label{Q2b}
Q(\alpha,2,3)(8\alpha^3)=\left[\begin{pmatrix}-\beta_2<\alpha\xi_1<\alpha\\
                          -\beta_2<\alpha\xi_2<\alpha\\
      -\alpha<\alpha\xi_3<\alpha\\
\end{pmatrix}\bigcup\begin{pmatrix}
-\beta_2<\alpha\xi_1<\alpha\\
-\beta_1<\alpha\xi_2<-\beta_2\\
-\alpha<\alpha\xi_3<\alpha\\
\end{pmatrix}\bigcup\begin{pmatrix}
-\beta_1<\alpha\xi_1<-\beta_2\\
-\beta_2<\alpha\xi_2<\alpha\\
-\alpha<\alpha\xi_3<\alpha\\
                        \end{pmatrix}\right]\\
                        \bigcap
\left[\begin{pmatrix}-\alpha<\alpha\xi_1<\alpha\\
-\beta_2<\alpha\xi_2<\alpha\\
-\beta_2<\alpha\xi_3<\alpha\\
\end{pmatrix}\bigcup\begin{pmatrix}
-\alpha<\alpha\xi_1<\alpha\\
-\beta_2<\alpha\xi_2<\alpha\\
-\beta_1<\alpha\xi_3<-\beta_2\\
\end{pmatrix}\bigcup\begin{pmatrix}
-\alpha<\alpha\xi_1<\alpha\\
-\beta_1<\alpha\xi_2<-\beta_2\\
-\beta_2<\alpha\xi_3<\alpha\\
\end{pmatrix}\right].
\end{multline}

Applying distributive property, the intersection of volumes is
computed, yielding:

\begin{multline}\label{Q2c}
Q(\alpha,2,3)(8\alpha^3)=\begin{pmatrix}-\beta_2<\alpha\xi_1<\alpha\\
                    -\beta_2<\alpha\xi_2<\alpha\\
-\beta_2<\alpha\xi_3<\alpha\\
\end{pmatrix}\bigcup\begin{pmatrix}
-\beta_2<\alpha\xi_1<\alpha\\
-\beta_2<\alpha\xi_2<\alpha\\
-\beta_1<\alpha\xi_3<-\beta_2\\
\end{pmatrix}\bigcup\begin{pmatrix}
-\beta_2<\alpha\xi_1<\alpha\\
-\beta_1<\alpha\xi_2<-\beta_2\\
-\beta_2<\alpha\xi_3<\alpha\\
                        \end{pmatrix}\\
                  \bigcup
\begin{pmatrix}-\beta_1<\alpha\xi_1<-\beta_2\\
-\beta_2<\alpha\xi_2<\alpha\\
-\beta_2<\alpha\xi_3<\alpha\\
\end{pmatrix}\bigcup\begin{pmatrix}
-\beta_1<\alpha\xi_1<-\beta_2\\
-\beta_2<\alpha\xi_2<\alpha\\
-\beta_1<\alpha\xi_3<-\beta_2\\
\end{pmatrix}.
\end{multline}

To find the $L=4$ hyper-volume, the surface in Eq. \ref{Q2a} should
be expanded to $4D$, all over $-\alpha < \alpha\xi_1<\alpha$,
$-\alpha < \alpha\xi_2<\alpha$ as follows:

\begin{multline}\label{Q2d}\left[\begin{pmatrix}-\alpha<\alpha\xi_1<\alpha\\
                          -\alpha<\alpha\xi_2<\alpha\\
     -\beta_2<\alpha\xi_3<\alpha\\
-\beta_2<\alpha\xi_4<\alpha\\
\end{pmatrix}\bigcup
\begin{pmatrix}
-\alpha<\alpha\xi_1<\alpha\\
-\alpha<\alpha\xi_2<\alpha\\
-\beta_2<\alpha\xi_3<\alpha\\
-\beta_1<\alpha\xi_4<-\beta_2\\
\end{pmatrix}\bigcup\begin{pmatrix}
-\alpha<\alpha\xi_1<\alpha\\
-\alpha<\alpha\xi_2<\alpha\\
-\beta_1<\alpha\xi_3<-\beta_2\\
-\beta_2<\alpha\xi_4<\alpha\\
\end{pmatrix}\right]
\end{multline}

and the volume in Eq. \ref{Q2c} should also be expanded to $4D$ all
over $-\alpha < \alpha\xi_4<\alpha$. Finally the intersection
between these $4D$ hyper-volumes must be found.

\begin{figure}
\includegraphics
[width=14 cm]{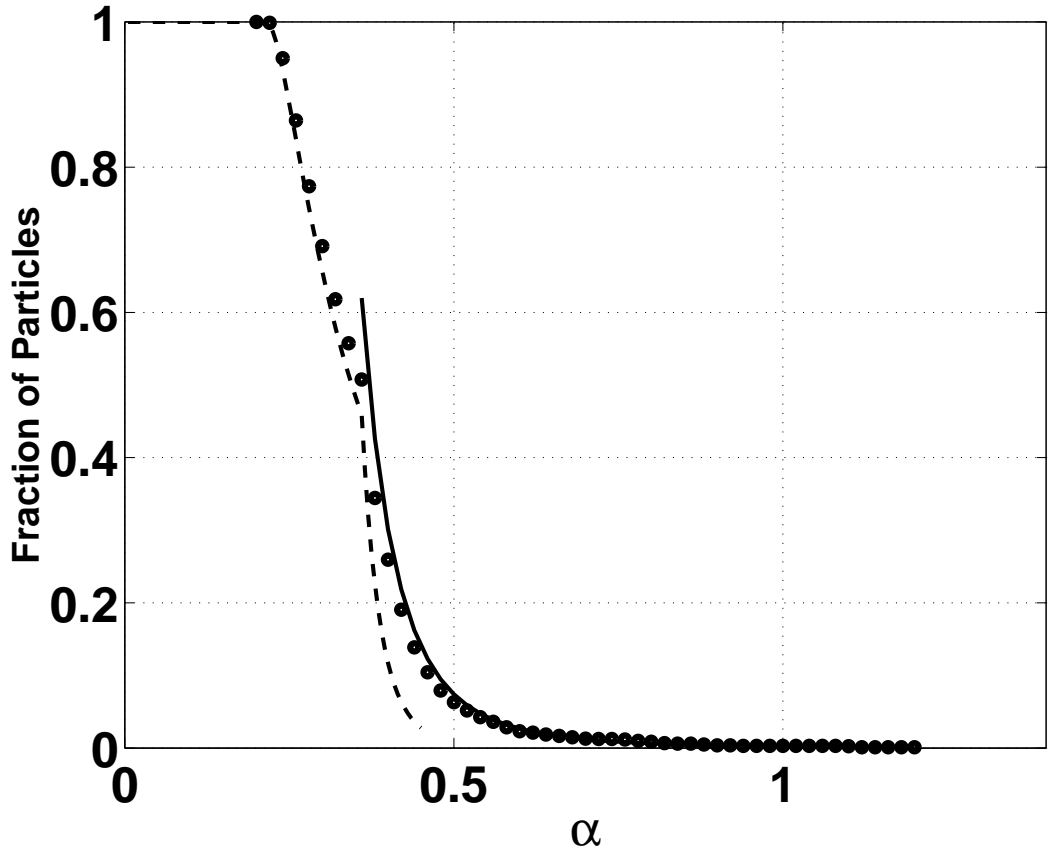} \caption { The fraction of particles
traversing a length $L=15$ disordered region is plotted as a
function of $\alpha$, for $\Gamma = 1.35$, $\omega=0.1$, $\gamma=1$
and $\mu=0.5$. Circle markers corresponds to simulations obtained
with $1600$ paticles. Dotted line corresponds to function
$Q(\alpha,K,L)$ in Eq. \ref{proba}, where no correlation is
considered between contiguous traps. Solid line corresponds to
$Q(\alpha,K,L)$ found by using CPM. }\label{Pdealfa}
\end{figure}

As $L$ grows, the number of component hyper-volumes in the set
increases and their intersections are hard to calculate by hand. We
calculated the resultant hyper-volume up to $L=15$ by using a binary
method in which intersections are 'and' boolean operators.  Fig.
\ref{Pdealfa} shows that there is excellent agreement when
$\alpha>\beta_{1}$ for $L=15$.

\section{Connection between Trapping and Transport
Properties}

\begin{figure}
\includegraphics [width=12 cm]{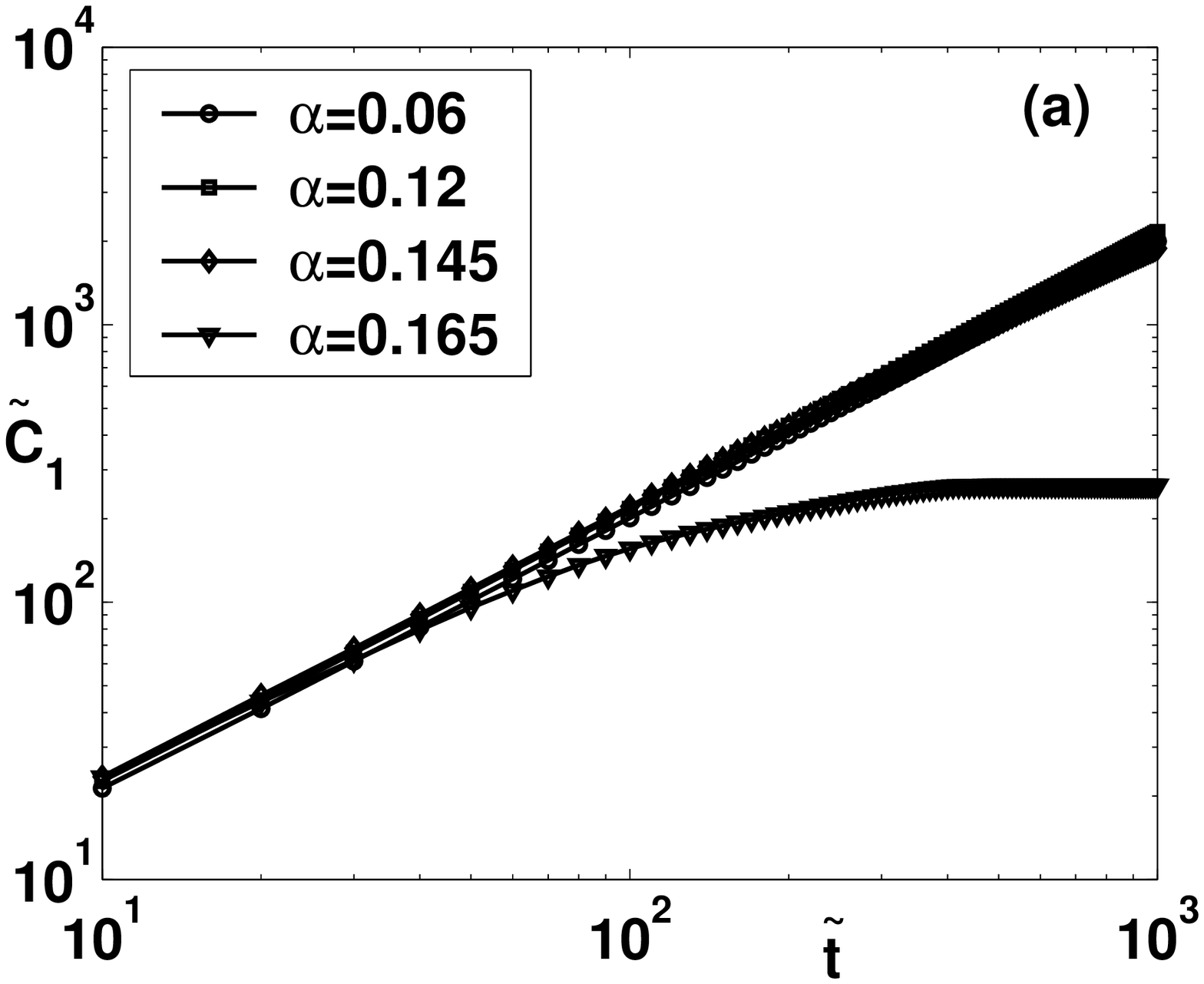}
\newpage
\includegraphics [width=12 cm]{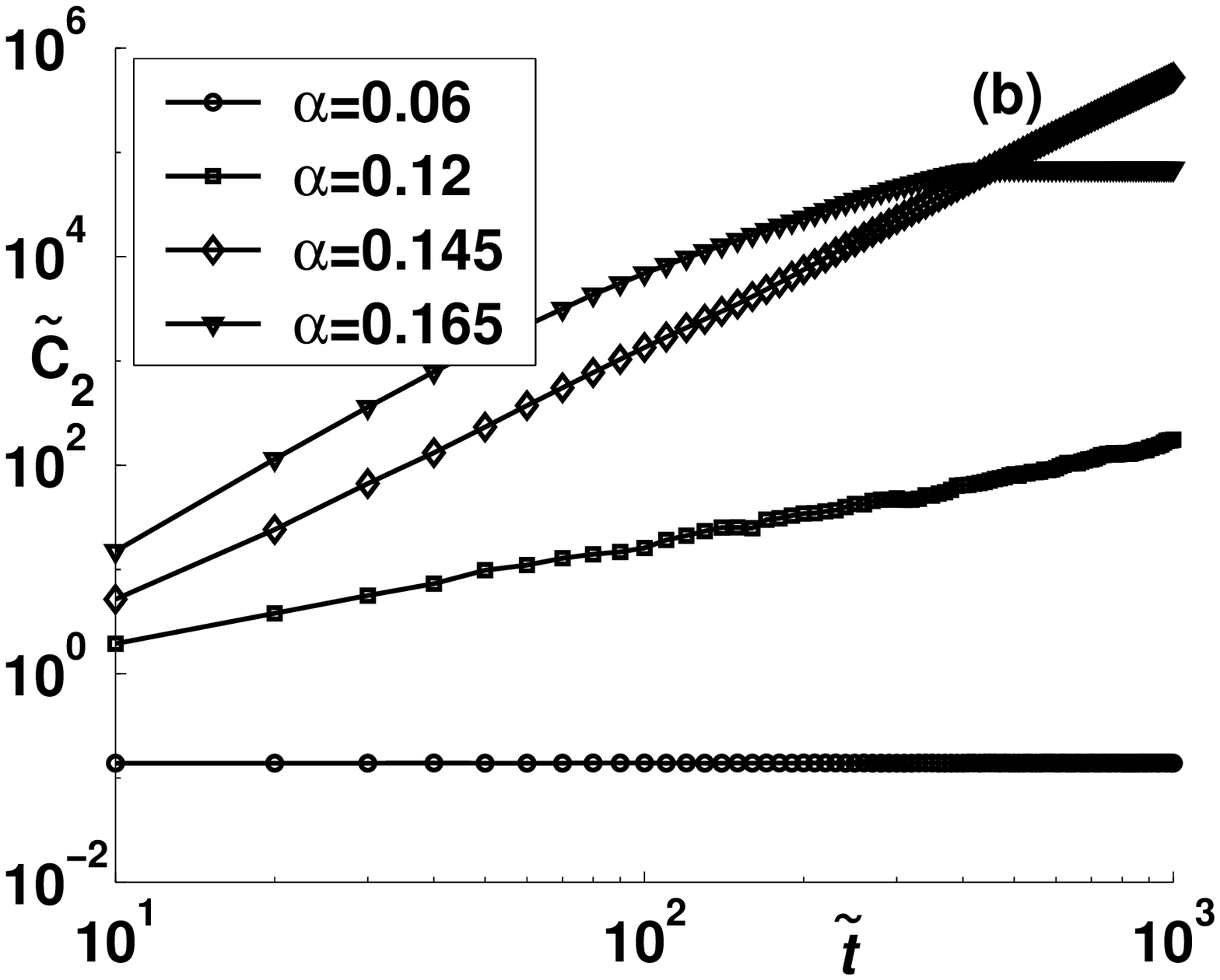} \caption { The cumulants  (a) $\widetilde{C_1}$  and
(b) $\widetilde{C_2}$ are plotted as a function of time
$\widetilde{t}$,  for $2200$ massless particles, traversing a $L=50$
disordered region, with $\Gamma = 1.65$, $\alpha=0.15$,
$\omega=0.1$, $\gamma=1$ and $\mu=0.5$.}\label{cum}
\end{figure}

\begin{figure}
\includegraphics
[width=8 cm]{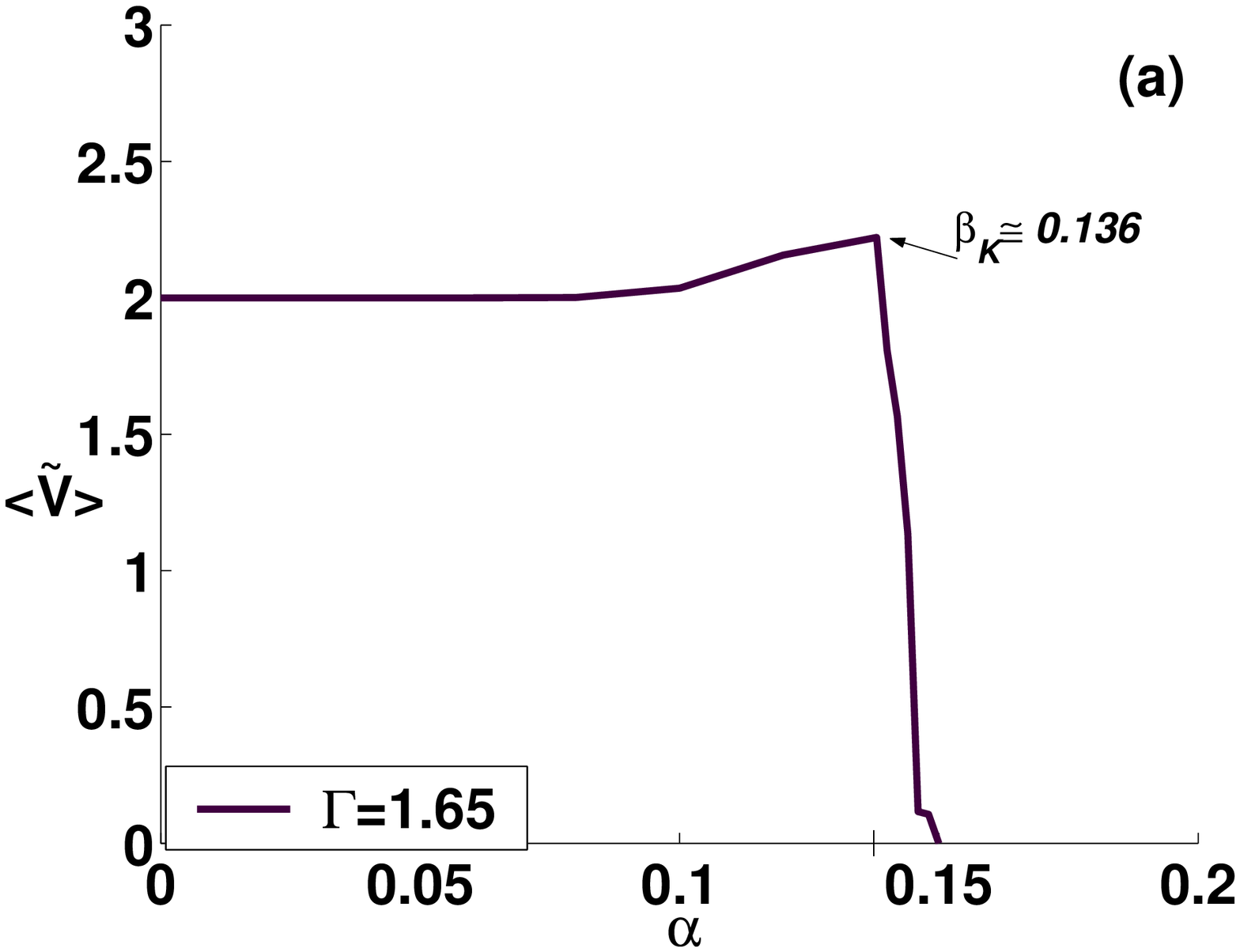}
\newpage
\includegraphics [width=8
cm]{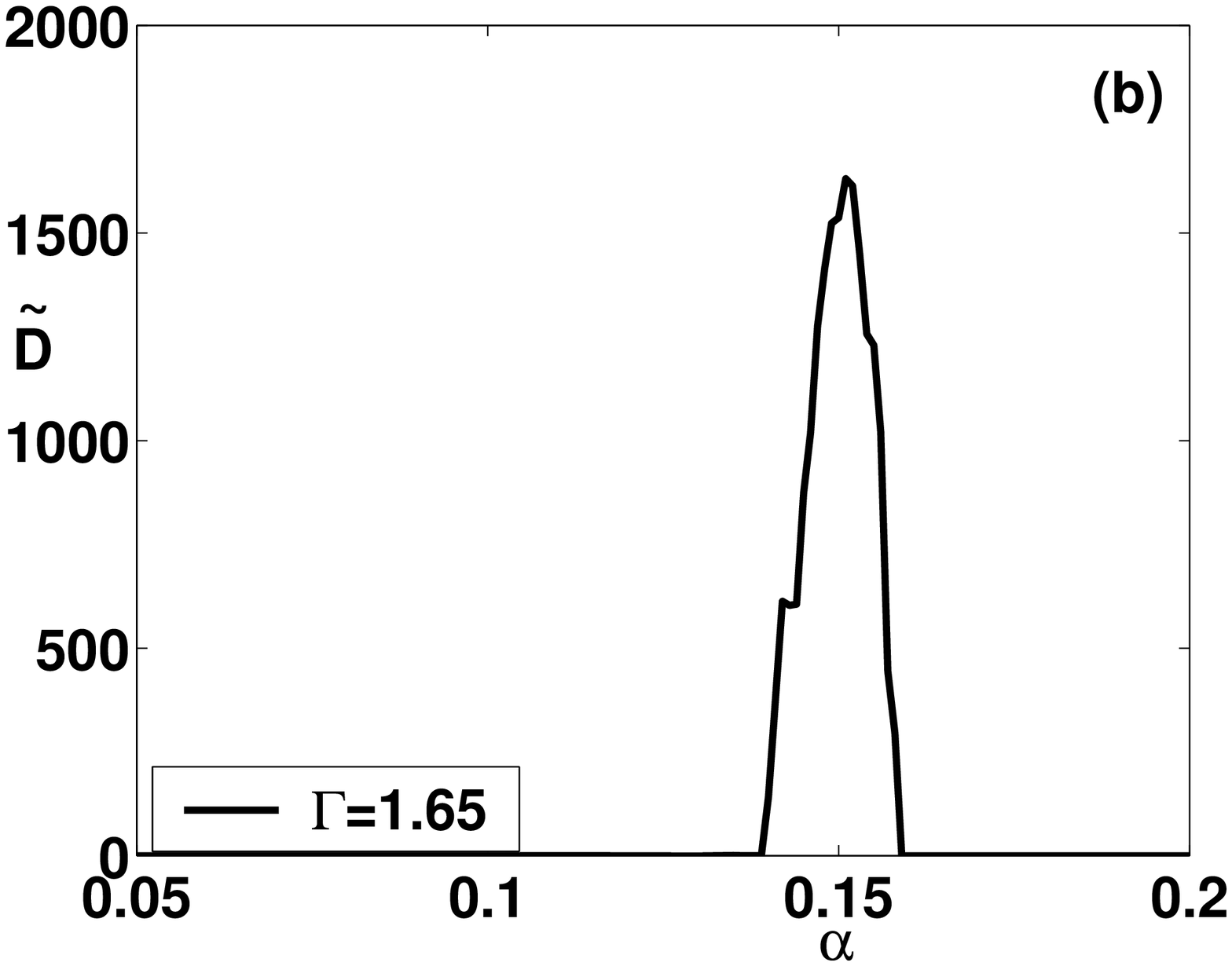}
\newpage
\includegraphics [width=8
cm]{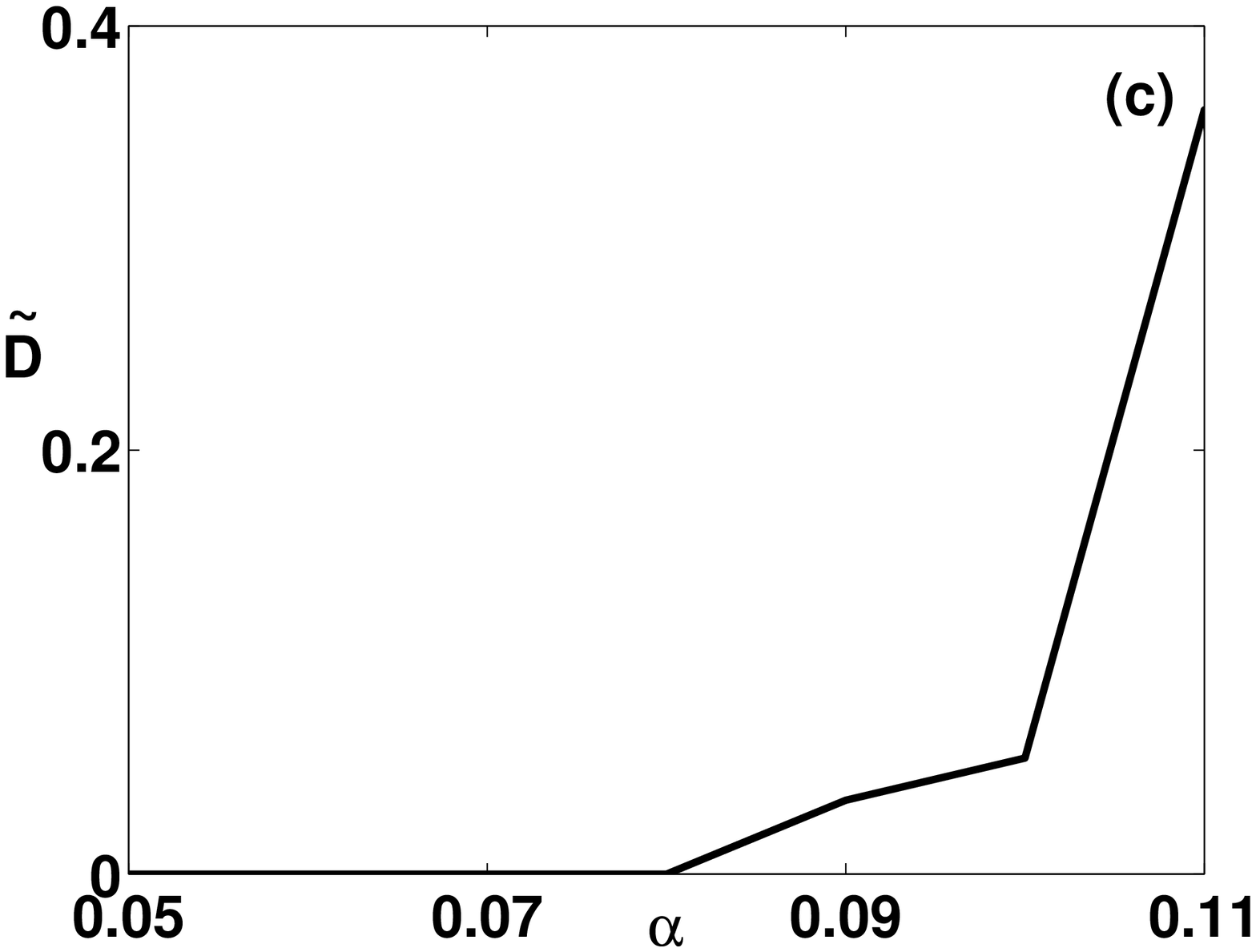} \caption {(a) The mean velocity $\langle
\widetilde{V}\rangle $ and (b),(c) the  diffusion coefficient
$\widetilde D$ are plotted as a function of $\alpha $ (for a packet
of $2200$ particles traversing a $L=50$ disordered region with
$\Gamma =1.65$, $\omega=0.1$, $\gamma=1$ and $\mu=0.5$).  (c) is an
enlargement of part of Fig. (b), where trapping does not exist and
the transport is diffusive}\label{vVSalfa}
\end{figure}

The presence of the traps has a significant macroscopic consequence
in multiparticle systems on the variation of the current
$\langle\widetilde{V} \rangle $ and the diffusion coefficient
$\widetilde{D}$.  In order to explain this consequence, consider the
cumulants $\widetilde{C}_{1}$ and $\widetilde{C}_{2}$, shown in Fig.
\ref{cum}, as a function of time for a packet of $2200$ massless
particles with $\Gamma=1.65$, $\omega=0.1$, $\gamma=1$, $L=50$ and
$\mu=0.5$. The corresponding derivatives $<\widetilde{V}>$ and
$<\widetilde{D}>$, are plotted as a function of $\alpha$ in Figs.
\ref{vVSalfa}a - \ref{vVSalfa}c.
   We note that as a function of $\alpha$, four regions with different
transport properties may be recognized in these figures:

1) Very low disorder strength region, $\alpha\leq0.08$. In this
region the disorder has no effect (see the case $\alpha=0.06$ in
Fig. \ref{cum}), the first cummulant $\widetilde{C}_{1}$ is
proportional to $\widetilde{t}$ and there is no diffusion as
evidenced by the constant value of $\widetilde{C}_2$ and the zero
value of $<\widetilde{D}>$ (see Fig. \ref{vVSalfa}c). Particles can
only reach the region $(+3,-1)$ of Fig. \ref{zonas} and the only
possible value of the mean velocity of the particles is $2$. Thus,
$<\widetilde{V}>$ remains constant equal to $2$ and the dynamics is
essentially the same as in the perfect lattice case.

2) Intermediate disorder region, $0.08<\alpha<\beta_K=0.136 $ (see
the case $\alpha=0.12$ in Fig. \ref{cum} and Fig. \ref{vVSalfa}c,
where this region is enlarged). In this region, the first cummulant
$\widetilde{C}_{1}$ is proportional to $\widetilde{t}$, but now
$\widetilde{C}_2\thicksim\widetilde{t}$, indicating the existence of
normal diffusion. The slope of the packet mean velocity changes
abruptly for each value of $\alpha$, where a new mode is reached, as
seen in Fig. \ref{zonas}.  At $\alpha>0.08$ the region labeled
$(+3,-0)$ is reached and the mean velocity may have one of the two
values $+2$ or $+3$. Consequently $\widetilde{V}$ increases with
$\alpha$, as can be seen in Fig. \ref{vVSalfa}a. At $\alpha=0.1$
region $(+4,-1)$ is also available and the mean velocity can have
any of the values $+2$, $+3$ and $+4$. Then $\widetilde{V}$
continues increasing with $\alpha$, with a higher slope. Fig.
\ref{vVSalfa}c shows an enlarged view of a region of Fig.
\ref{vVSalfa}b, where the normal diffusion can be observed.  As can
be seen, for  $\alpha<0.08$ transport is not diffusive. At
$\alpha=0.08$ normal diffusion starts and as $\alpha$ is increased
further, diffusion is enhanced and the same critical value of
$\alpha$ appear as in the case of $\widetilde{V}$
(Fig.\ref{vVSalfa}a).

3) The region $\alpha\simeq\beta_K=0.136$ (see, for example, the
case $\alpha=0.145$ in Fig.\ref{cum}).  In this region the trapping
mechanism has already started. The cumulants $C_1$ and $C_2$
increase as $\widetilde{t}^H$ with $H>1$, indicating a
super-diffusive behavior. The time-dependence of the second moment
is very complicated during the trapping process and it strongly
depends on both the quenched disorder realization and the strength
of the disorder. This transitory time is considerably reduced as
$\alpha$ increases beyond the threshold. The trapping mechanism
produces an abrupt descent in $\widetilde{V}$. The other values
$\beta_{K-1}=\beta_2 $ and $\beta_{K-2}=\beta_1$ do not appear in
Figs.\ref{vVSalfa} because the probability that a particle gets
trapped in a $K=3$ valley length trap approaches $1$ for a disorder
region with $L=50>>K=3$, and most particles get trapped in $K=3$
length valleys, regardless of whether $K-1=2$ and $K-2=1$ length
traps exist or not. In Fig. \ref{vVSalfa}b the super diffusive
region is clearly recognized by the high values in $\widetilde{D}$.
Below the threshold value $\beta_K$ diffusion is normal and
$\widetilde{D}$ is much smaller than in the super diffusive region.
Note that $\widetilde{C_2}$ varies with time in a very complicated
way as long as there exist both, trapped and untrapped particles.
Untrapped particles suffer a normal diffusion process but trapped
particles cause the packet to get wider as it evolves, increasing
$D$. In fact, for $t\rightarrow\infty$ (and consequently
$L\rightarrow\infty$) all particles get finally trapped and
$D\rightarrow 0$ .

4) The region $\alpha\gg\beta_K$ (see the case $\alpha=0.165$ in
Fig.\ref{cum}).  In this region, complete trapping occurs even for
small disordered zones and both $<\widetilde{V}>$ and
$\widetilde{D}$ decrease to zero.
%
%
%%%%%%%%% begin of changed paragraph%%%%%%% %%%%%%% %%%%%%% %%%%%%% %%%%%%% %%%%%%% %%%%%%%
The system undergoes a transition at $\alpha=\beta_K$ between two
different transport regimes: for $\alpha<\beta_K$ there is normal
diffusive transport while for $\alpha>>\beta_K$ both transport and
diffusion disappear. In the neighborhood of $\alpha=\beta_K$
anomalous diffusion is present. The anomalous transport found
recently in an overdamped tilted potential model with thermal noise
but without quenched disorder and driving force \cite{reimann2}, a
similar transition between normal diffusive transport is given by
the variation of the tilting force $F$ and anomalous diffusion
appears near the critical value $F_c$.

The anomalous transport effect of trapping is robust under thermal
fluctuations in the sense that thermal fluctuation amplitudes of the
order of the quenched disorder strength are required to destroy this
effect.
%%%%%%% end of changed paragraph%%%%%%% %%%%%%% %%%%%%% %%%%%%% %%%%%%% %%%%%%%
%
%
\section{Conclusions}

A novel trapping mechanism  is discovered which is proposed as the
origin of the anomalous transport in a multi-particle overdamped
disordered ratchet. It is found that once a particle reaches a trap
it remains localized inside a small region, oscillating
synchronously with the external force. By means of the disorder
configuration space, critical values for the disorder strengths were
determined and the fraction of particles traversing a disordered
region were obtained by means of two methods. In the first method,
valid for low disorder, the effects of correlations between the
contiguous traps were neglected. In the second approach, which is
valid for disorder strengths over all the critical values,
correlations were considered. The probability density function shows
excellent agreement with the simulation data. The trapping mechanism
presented here explains the singular behavior of velocity and
diffusion with disorder in overdamped ratchets reported in
\cite{popescu}. This analysis may be helpful in the study of other
systems exhibiting strange kinetics and also for practical
applications such as the design of particle separation techniques in
multi-particle systems.

\section{Acknowledgments}
This work was partially supported by CONICET (PIP 5569), Universidad
Nacional de Mar del Plata and ANPCyT (PICT 11-21409 and PICTO
11-495).

\newpage

\newpage

\newpage

\newpage

\newpage

\newpage

\end{document}